# The Anatomy of a Modular System for Media Content Analysis

Ilias Flaounas, Thomas Lansdall-Welfare, Panagiota Antonakaki, Nello Cristianini

Intelligent Systems Laboratory, University of Bristol, Bristol BS8 1UB, United Kingdom

Intelligent systems for the annotation of media content are increasingly being used for the automation of parts of social science research. In this domain the problem of integrating various Artificial Intelligence (AI) algorithms into a single intelligent system arises spontaneously. As part of our ongoing effort in automating media content analysis for the social sciences, we have built a modular system by combining multiple AI modules into a flexible framework in which they can cooperate in complex tasks. Our system combines data gathering, machine translation, topic classification, extraction and annotation of entities and social networks, as well as many other tasks that have been perfected over the past years of AI research. Over the last few years, it has allowed us to realise a series of scientific studies over a vast range of applications including comparative studies between news outlets and media content in different countries, modelling of user preferences, and monitoring public mood. The framework is flexible and allows the design and implementation of modular agents, where simple modules cooperate in the annotation of a large dataset without central coordination.

## 1 Introduction

The ready availability of vast amounts of digital data and the creation of new powerful methods of analysis have started transforming many branches of science, opening the possibility for data-driven approaches and science automation to fields as diverse as biology, physics, chemistry and even the humanities and social sciences [1, 2]. In the social sciences, one area of particularly intense progress is the study of traditional news media and also the analysis of the new social media. Both of these applications generate large quantities of readily available data for media analysts and social scientists in general to process and investigate.

Traditional research in this area has relied on the labour intensive step of human coding: the activity of human experts reading and annotating news or other media items. Recent works have shown that Artificial Intelligence (AI) algorithms can be deployed to automate many steps of this expensive process, therefore paving the way to the analysis of much larger sets of data. This automation has become possible because of the convergent effect of two trends: the availability of data and the technology to manage it; and the emergence of a new generation of powerful (mostly statistically-driven) algorithms for machine learning, data mining and text analysis.

One important observation is key when combining recent advances in AI algorithms with the recent trend to automate scientific research in social sciences: the basic tasks that are solved by the classical AI algorithms do not directly coincide with the tasks that are of value to social scientists. In other words, basic social science tasks such as measurement of gender bias in a text are the result of multiple elementary AI algorithms, such as the detection of named entities, co-reference resolution, topic detection and so on. This situation directly highlights the main technological challenge presented by automation of science in general, and of social sciences in the present case: the design of integrated intelligent systems that allow many AI algorithms to collaborate to extract value from data. Namely, a framework for the combination of multiple AI algorithms that is principled and "independent" of the specific algorithms used for the annotation.

The design and implementation of modular systems is a general challenge for AI, where significant progress has been achieved in the optimization of single-task algorithms such as classification, but where most interesting applications call for complex tasks that require the coordinated usage of many such methods. The understanding of how complex tasks can be decomposed into modules is an important part of modern AI, as well as the inverse problem of integrating simple modules to generate complex behaviour.

In this paper we describe a solution for the design of large scale intelligent systems that combine multiple AI modules, and that allows the automatic annotation of large amounts of data for scientific purposes. These modules are combined in a coherent and scalable framework which we describe. Besides describing the system we have developed over the past few years for our own research, the method we present is general and likely to be useful in many other domains, as discussed in Section 5. All communication between modules in our approach is obtained by reading and writing on a shared blackboard, and decisions are made by the system without any centralised control. Intelligent goal driven behaviour emerges in our system as the result of the interaction of all these modules, calling each other, without central control. Of course the system is designed, but the behaviour is emergent.

## 2 System Overview

Typically simple computational analysis systems found in AI literature are single purposed and usually implement a specific pipeline-based architecture: Data flow in a linear way from one analysis component to the next, and some specific output is produced at the final step. Our system is designed to be highly modular and each module performs a specific analysis task. It is designed to achieve a series of key properties including flexibility, robustness and no centralised decision making. Flexibility means that the processing lines are not hardwired but that they can easily change and modules can be added or removed without interrupting or needing to adapt current modules. Robustness is the property where we want the system to produce the most accurate results possible, even if a specific module fails. In a simple pipeline the failure of a module will break the processing line. Robustness is also related with the notion of no single point of failure. That means that the system should be able to handle hardware or network failures that are expected to happen in the long run. The no centralised decision making means that we want to avoid having a meta-module that will organise the behaviour of the existing modules. Some background on the ideas that affected our system design are provided in Section 5.

An overview of the system is shown in Fig. 1. Modules are organised in three groups: input, output and analysis. Data live in a database and are processed by the modules through a common Application Programming Interface (API). The input modules create new data by crawling the web or, in the case of machine translation, by translating news articles into English; the analysis modules analyse the data by adding annotations to each item; and the output modules create reports or populate websites that provide insight to the final user.

At the logical level the data are organised in a series of blackboards as illustrated in Fig. 2. Each blackboard contains similar items like, for example, news articles or tweets. The items are accessible by modules which can alter them by adding or removing annotations. The modules do not communicate directly with each other, but only indirectly through the blackboards. The background literature we used as a basis for the developed architecture is discussed in the last section of the paper.

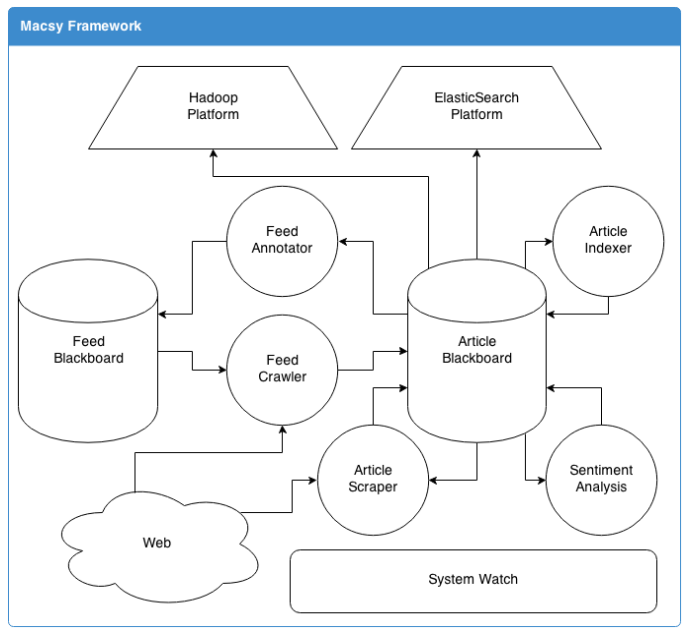

*Fig 1. Simplified overview of the Macsy framework, demonstrating the interaction of modules (circles) to annotate data via central blackboards.*

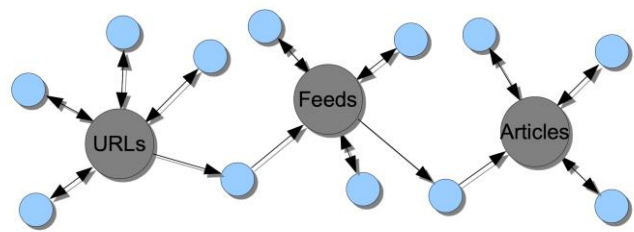

*Fig. 2. Data is organised at the logical level in blackboards (grey), and are annotated by modules (blue).*

### 2.1 Blackboards

In our system, a blackboard is a shared repository of data, and all data is organised in a set of blackboards. There are two main blackboards, one for storing news articles, and one for storing tweets. For the news analysis, we have two more blackboards, one for storing news feeds that are the sources of news stories, and one for storing outlets. Each outlet corresponds to an individual publisher, such as newspapers, broadcast stations or blogs. We use website domain names to define what constitutes a publisher. One outlet can have one or more news feeds, and each article can come from one or more news feeds. In a similar way a different blackboard holds the queries used to query the location of 54 cities of UK in Twitter. Other auxiliary blackboards include the Locations blackboard that holds the geographic position of cities, geographic regions or other places of interest and the URLs blackboard that is used by the FeedFinder module to explore the web in order to find novel news sources.

Each blackboard holds its own set of tags, i.e. textual strings, for annotating the data. Tags are important for a number of reasons: they define overlapping sets of items, e.g. the topics covered in an article, they convey information like the language an article is written to, and they define the modules that will operate on the item as we discuss in the following section.

## 2.2 Modules

Modules are organised in three categories: Input modules that create the basic content that populates blackboards; Analysis modules that annotate the blackboards' content and; Output modules that provide the output of the system.

The input modules include the news crawler and the Twitter crawler. The first crawls the web for the items in the feeds blackboard and returns their content as individual articles in the articles blackboard. For each article the crawler provides the title, a short description, the publication date, and a link to the HTML page that contains the full content of the article. The articles are tagged with the language that they are written (if known), the location of the outlet that published them, and the ID of the feed that carried them. Also, the same module generates a hash based on title, description and outlet ID that is used to identify identical articles. This is very useful since it allows the module to quickly identify duplicate articles that should not be re-added to the blackboard. This happens when the crawler is relaunched and finds the same content again from the same sources; or when the same article is published in more than one feed. In the later case we add any extra information that the second feed provides to the article. Similar functionality is implemented in the Twitter crawler, which collects tweets from a predefined list of locations. Currently we monitor and collect data from 54 major UK cities and we collect 500K tweets per day.

Modules are launched with some predefined frequency, for example every hour, or once per day. A module starts its operation by querying a blackboard in order to find a set of items that contain some specific set of tags and/or annotations. For example a topic tagger will search for items that have a special tag "FOR>SportsTagger". That tag is placed on items by some different module like the feature extractor. An upper limit of items processed per module-run in order to bound the module running time. The module processes each item, one after the other in an independent way. Of course, a module can implement a multithreaded approach so that multiple items can be processed in parallel. The result of the analysis is a set of new tags and annotations. The topic tagger for sports will annotate the relevant items with the tag "Sports". Also, it has the option to add extra tags that may be used to trigger the launch of another module, like a module that creates reports based on the sports articles of the day.

Modules can store their own private information outside a blackboard but this information is not shared with other modules. For example, a module may store a data model that it uses to make predictions. If two modules need to communicate, i.e. exchange information, this can be achieved only by reading and writing the information on the items of a blackboard.

A common issue when building a modular system is deciding on the appropriate size of a module. At the one end a small module can only perform some very trivial processing, while at the other end a large module can perform an overly complex task which could be divided in smaller parts. In our architecture we follow the idea that a module should be large enough to be able to create specific annotations to items that are useful to another module. Next, we summarise some of the basic implemented modules that constitute our system:

- **HTML Scraper**. The scraper parses the HTML page that contains the full content of an article and returns the raw textual information of the article after removing irrelevant text from the page, along with images, menu items and any HTML code.

- **Feed Finder.** This module implements a focused crawler that searches the web in order to find novel news sources in RSS or Atom format. Currently, the addition of news feeds to the watch list is semi-automatic since human approval is needed.

- **Machine Translation.** Currently this module is able to machine-translate 21 European languages into English. The module is based on the popular Phase Based Model statistical machine translation approach. Every non-English article is tagged to be fed into machine translation and the result is written in the database as a new machine-translated article. We based the module implementation on the open source software Moses.

- **Feature Extractor.** It creates a vector representation of the article based on TF/IDF features. The module implements a typical text processing pipeline including stop-words removal and stemming. The output is used by a

- series of analysis modules like the mood or topics classifiers.

- **Language Detector.** It annotates an article based on the language used.

- **Sentiment Extractor.** The module measures and annotates articles based on whether they contain adjectives that carry sentiment.

- **Mood Detector.** The module computes the inner product of each article with a list of predefined mood-related words and it annotates items with a score. We track four moods: joy, anger, fear and sadness.

- **Topic Detector.** The module annotates articles based on an SVM classifier trained on standard datasets like the Reuters and The New York Times corpora. Topics we track include politics, business, sports, crime, war and religion.

- **Geolocator**. The module identifies the mention of locations in articles. It also deploys some location disambiguation algorithms to identify the correct location between locations that have the same name.

- **Readability Annotator.** Provides scores based on how readable a document is. The score is based on the FLES readability score, and it is based on simple metrics like average length of sentences and average number of syllables per word.

- **Popularity Annotator.** This module measures the popularity of an article. The popularity is based on a linear model built using an online ranking algorithm and data from specific outlets that publish a list of their most popular articles. This algorithm is trained by comparing the articles that managed to become popular versus the articles that did not become popular although they were published by the same outlet and on the same day.

Some of the modules were developed in house, while others act like wrappers of existing NLP or machine learning libraries. For example, the topic classifiers are based on the LibSVM library while the machine translation is based on the Moses library.

Finally, the output modules typically create reports or some XML files that can be used to populate the content of demo websites. In Sect. 4 we present some exemplary case studies that show the capabilities of the output modules.

## 3 Implementation and Data Management

Modules are implemented in Java and are typically comprised of an executable and a settings file. All modules use the same API that was developed in order to guarantee the homogeneous use of blackboards across the system. The API also serves as an intermediate layer between the database and the modules allowing the change of the underlying database management system without the need of changing the modules' code. The settings file typically includes the module name; a short description of the module functionality; the name of the blackboard that is used as input; the set of tags and fields that items should have in order to be processed by the module; the name of the blackboard that is used as output (usually it is the same as input); and the name of the tags and annotations that the module will add to the items. The settings file also define the maximum number of items that the module will process in each launch and the number of threads that it is allowed to initiate (modules are typically multithreading for performance reasons). This allows the system to have multiple modules with similar functionality without the need of having more than one binary executable. For example, the topic detectors are all based on the same executable but each one has a different settings file.

Modules are replicated and distributed across multiple physical machines for additional robustness. Currently our system is organised on eight physical machines. Modules are allocated specific time slots in machines, i.e. they are triggered in predefined times and they have an upper timeout limit enforced by operating system. This guarantees that no module can abuse computational resources. Also, a web interface we call SystemWatch has been developed, that allows some basic administration tasks and the monitoring of the system performance. For example, the interface allows observation of the status of the physical machines, and the monitoring of modules status' and their input/output, such as how many articles were collected and how many were analysed per module.

Storing and management of data is an important factor in our system. The first version of the system was built around a MySQL database [3]. Recently the database system was replaced by MongoDB, a modern NoSQL solution. This database management system has a number of benefits that makes it quite attractive. First of all it is a schema-less document based database which allows enormous flexibility: all annotations of an item are stored with the item in the same table. Thus, a single query may return all the annotations about an item, without the need of having a central place to store which tables are available for which items. Also, the database is inherently distributed allowing reading and writing from multiple machines. This not only improves the performance of the database but it also highly increases the availability of the system since there is no single point failure. Currently our database is organised on four physical machines. The machines are logically organised in two pairs (shards). The data is split between the two shards and each shard is comprised of two machines that store the same data for extra availability and performance. Note that machines are physically located in two different buildings for avoiding external problems like power or network availability.

## 4  Case Studies – Application Scenarios

Our system has been used successfully in various social science projects including the analysis of Twitter content [4], the analysis of traditional news media [5], the analysis of the EU mediasphere [6], etc. Here, we present two case studies as representative results of the current functionality of the system, one on the comparison of news outlets and one on the sentiment analysis of Twitter content.

### 4.1 Comparison of News Outlets on Topics, Writing Style and Gender Bias

A popular topic of research in media studies is the detection of differences or biases among news outlets. We showed how similar studies can be performed using our system in our previous work [5]. We analysed a large corpus comprised of 2.5 million news articles collected from 498 news outlets over a period of 10 months. For each article we identified the general topics that it covered, as well as two basic writing style properties, namely the readability and linguistic subjectivity.

The computation of the aforementioned quantities allowed the answering of a series of research questions. For example, for the articles of each topic we calculated the average readability, finding that articles about sports are the easiest to read while articles on politics are the hardest to read; and for linguistic subjectivity, finding that articles about fashion and arts are the most linguistically subjective, while business articles were the most objective. Furthermore, we directly compared 15 major US and UK newspapers on which topics they tend to cover more often and their writing style. In Fig. 3 we visualise the comparison of outlets based on their writing style: outlets with similar writing style are closer together.

### 4.2 Sentiment Analysis of Twitter Content

Measuring the current public mood is a challenging task. The traditional approach would require questioning a large number of people about their feelings. However, social media, such as Twitter, can easily become a valuable source of information about the public due to the fact that people use them to express their feelings in public.

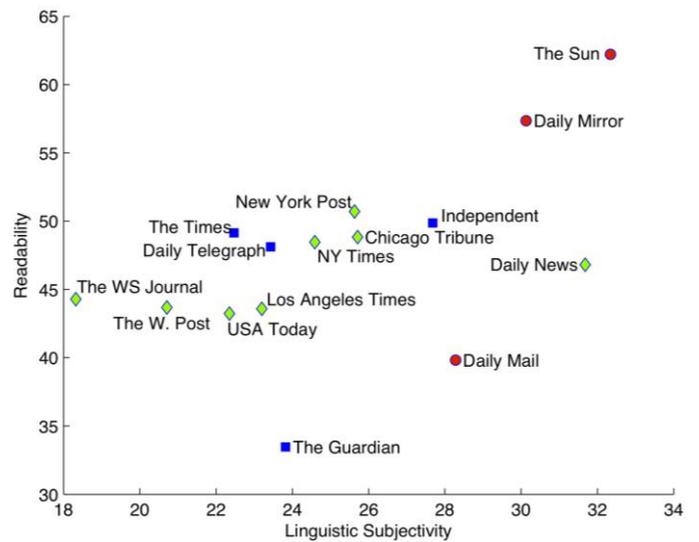

*Fig. 3. Comparison of news outlets in the US and UK based upon their writing style.*

As demonstrated in our study [7], it is feasible to capture the public mood by monitoring the stream of Twitter data. The dataset that was analysed was comprised of 484 million tweets that were generated by more than 9.8 million users in UK, between July 2009 and January 2012. We focused on tracking four moods and for each mood we generate one timeline of the volume of related tweets. The further analysis of these timelines reveals that each of the four emotions changes over time in a rather predictable manner. For example, we found a periodic peak of joy around Christmas and a periodic peak of fear around Halloween. More surprisingly, we found that negative moods started to dominate the Twitter content after the announcement of massive cuts in public spending on October 2010. In Fig. 4, we plot the mood levels for the period of study and we visualise them as a facial expression using the Grimace tool.

## 5  Discussion

The architecture we have developed for the creation of our media-content analysis platform presents a general solution to the problem of integrating very diverse algorithms into a single modular system. We anticipate that this approach can be applied to many other scenarios where data needs to be processed in a collaborative fashion by multiple software modules.

One of its most appealing features is the way in which the various modules cooperate to annotate the data, in a decentralised manner: communication between modules takes place only via annotation left by them on the items contained in the common blackboard. This is an instance of stigmergic communication. Stigmergy is a mechanism for indirect coordination between agents, obtained by each agent leaving traces (stigmata) in the environment while performing an action, that can be read by another agent, and affect its behaviour, for example triggering another action. In nature this mechanism is used to generate coherent collective behaviour, and is one of the ways in which complex systems self-organise without need for central planning or control, or for direct communication between individual agents. This mechanism allows efficient collaboration to take place between very simple agents who lack memory, planning and other cognitive capabilities. It is the way in which ants coordinate their behaviour, for example, the same method is found across many natural and artificial systems [8].

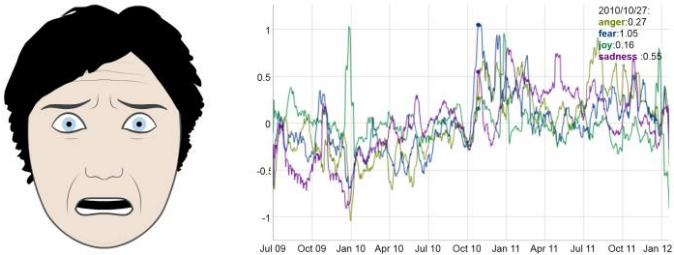

*Fig. 4. Visualising mood levels in UK Twitter over time with facial expressions.*

We implemented the general approach to distributed, decentralised coordination in a modular system by using a blackboard architecture, whereby all communications between modules are forced to take place via the reading and writing on the common shared information (blackboard). This provides a simple solution to the notoriously difficult problem of segmenting complex behaviour into several simpler modules: by insisting that only limited information is passed among modules, via the blackboard, we find a natural way to decompose complex processing into modules. The idea of using a common blackboard where multiple agents can read and write is very old in Artificial Intelligence, going back at least to the influential Pandemonium system created by Oliver Selfridge in 1959 to coordinate the action of several daemons [9].

It is also worth noting that our approach to modularity is consistent with the basic axioms listed by Fodor [10, 11] to be expected in modular systems: Domain specificity, i.e., modules are specialised to operate on specific kinds of input; information encapsulation, i.e., modules do not need information from within other modules to operate, they only process the input they are provided using their own private information; mandatory operation, i.e., modules process all their inputs without choice; and shallow outputs, i.e., they produce simple outputs, in our case typically 'tags'. The last two of his axioms are relative to biological systems and therefore are omitted here. Our approach to modularity is also related to the class of agent architectures known as "reactive robotics" described by Brooks [12].

In the literature there is a plethora of works that focus on extracting information from public news datasets using some specific methodology or tool. On the other hand, there are very few computational systems and approaches that are oriented to the automation of more than a single perspective of news analysis. Some interesting examples include the Europe Media Monitor where they provide a coherent summary of current news [13]; and the work by Castillo et al. for the analysis of television news programs [14].

Currently in our group work is under way for the creation of a computer vision system based on the same principles, where multiple modules cooperate in the extraction of information from large quantities of images collected from news sites. Finally, we want to note that core parts of the system are distributed as an open source library that we call Macsy ("Modular Architecture for Cognitive Systems"), at Github: https://github.com/mediapatterns/Macsy. The library contains the API that is used to coordinate the modules and keep track of the blackboards, as well as a series of implemented modules. It is built on top of MongoDB and it is developed in Java.

## 7 Acknowledgments

This research is funded by European Community's Seventh Framework Programme under grant agreement N° 270327 (CompLACS Project)